\documentclass[10pt,letterpaper]{article}
\usepackage[top=1in,left=1in,footskip=1in,marginparwidth=1in]{geometry}

\usepackage[utf8]{inputenc}
\usepackage{subfig}
\usepackage{enumitem}
\usepackage{dirtytalk}

\usepackage{natbib}
\setcitestyle{authoryear,open={((},close={))}} 

\usepackage{nameref,hyperref}

\usepackage[right]{lineno}

\usepackage{microtype}
\DisableLigatures[f]{encoding = *, family = * }

\usepackage{changepage}

\usepackage[aboveskip=1pt,labelfont=bf,labelsep=period,singlelinecheck=off]{caption}

\makeatletter
\renewcommand{\@biblabel}[1]{\quad#1.}
\makeatother

\usepackage{lastpage,fancyhdr,graphicx}
\usepackage{epstopdf}
\fancyheadoffset[L]{2.25in}
\fancyfootoffset[L]{2.25in}

\usepackage{color}

\definecolor{Gray}{gray}{.25}

\usepackage{graphicx}

\usepackage{sidecap}

\usepackage{wrapfig}
\usepackage[pscoord]{eso-pic}
\usepackage[fulladjust]{marginnote}
\reversemarginpar

\begin{document}
\vspace*{0.35in}

\begin{flushleft}
{\Large
\textbf\newline{CelluloTactix: Empowering Collaborative Online Learning through Tangible Haptic Interaction with Cellulo Robots.}
}
\newline
\\
Hasaru Kariyawasam \textsuperscript{1},
Wafa Johal \textsuperscript{2,*},
\\
\bigskip
\bf{1} UNSW, Sydney, Australia
\\
\bf{2} University of Melbourne, Australia
\\
\bigskip
* wafa.johal@unimelb.edu.au

\end{flushleft}

\section*{Abstract}
Online learning has soared in popularity in the educational landscape of COVID-19 and carries the benefits of increased flexibility and access to far-away training resources. However, it also restricts communication between peers and teachers, limits physical interactions and confines learning to the computer screen and keyboard. In this project, we designed a novel way to engage students in collaborative online learning by using haptic-enabled tangible robots, Cellulo. We built a library which connects two robots remotely for a learning activity based around the structure of a biological cell. To discover how separate modes of haptic feedback might differentially affect collaboration, two modes of haptic force-feedback were implemented (\textbf{haptic co-location} and \textbf{haptic consensus}). 
With a case study, we found that the haptic co-location mode seemed to stimulate collectivist behaviour to a greater extent than the haptic consensus mode, which was associated with individualism and less interaction. While the haptic co-location mode seemed to encourage information pooling, participants using the haptic consensus mode tended to focus more on technical co-ordination. This work introduces a novel system that can provide interesting insights on how to integrate haptic feedback into collaborative remote learning activities in future.

\begin{figure}[h]
  \includegraphics[width=\textwidth]{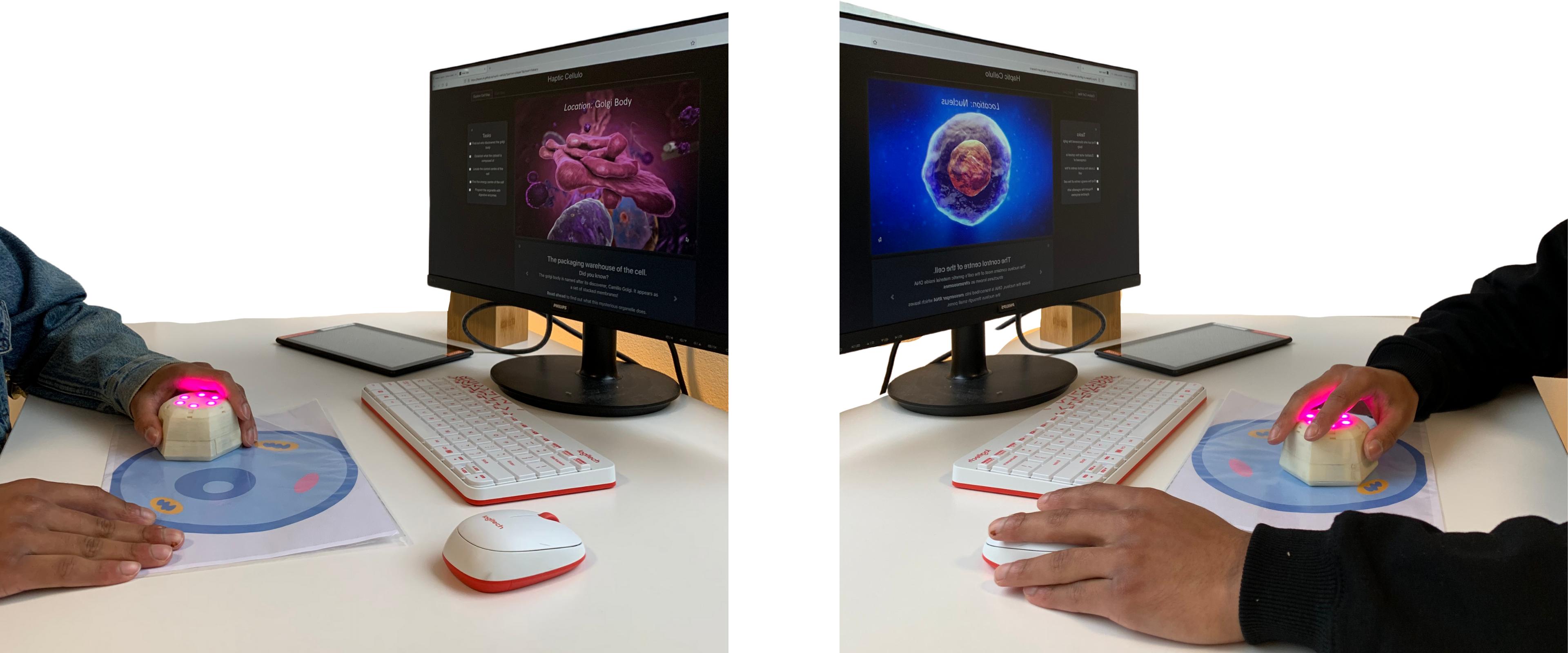}
  \caption{Remote Haptic Collaboration Setup featuring the printed paper sheet, the Cellulo robot used as a probe to explore the cell's organelles.}
  \label{fig1}
\end{figure}

\section*{Introduction}
Cellulo robots \cite{ozgur_cellulo_2017} are small, low-cost tangible robots which have been designed for collaborative use in the classroom. Each robot is capable of autonomous movement and can therefore interact with the user or with other Cellulo robots, displaying swarm-like behaviour. During learning activities, the robots are placed on a sheet of paper printed with a microdot pattern used for self-localisation \cite{ozgur_cellulo_2017}. Six LED lights are positioned on top of each robot, allowing it to display RGB colours. They are haptic-enabled and can elicit planar force-feedback. In a departure from other haptic-enabled technologies, the robots feature a haptics controller which allows them to simultaneously move and be moved by users \cite{ozgur_cellulo_2017}. This allows them to exert force on the learner's hand while the learner is using them, thus generating new opportunities for creating modalities of interaction with tangible interfaces.
The design motivation of Cellulo robots is to function as ubiquitous tools of the classroom, eventually becoming the pen and paper of the future. Whereas social robots tend to share humanoid features and take on a social character \cite{belpaeme_social_2018}, Cellulo robots are instead positioned as practical and readily replaceable tools for teaching \cite{ozgur_cellulo_2017}. Ease of replaceability of each robot is achieved through a distinct lack of personalisation. The robots do not invoke the formation of affective attachment, so users can interchange them freely. Rather than encouraging social bonding, this class of tangible robots aim to embody digital information in a physical form \cite{east_actibles_2016} which can then be manipulated by human individuals at will. The experience of interacting with Cellulo is likened to interacting with virtual points sitting on a plane \cite{ozgur_haptic-enabled_2017}, which can have a variety of applications in the classroom. One activity demonstrated in \cite{ozgur_cellulo_2017} is a virtual treasure hunt, where students would scan regions of a treasure map printed on paper using a Cellulo robot, and then \emph{pull} treasure out of the map by turning the robots. Cellulo here plays the role of a passive robot which reacts to the environment in which it is localised. Alternatively, Ozgur et al. \cite{ozgur_cellulo_2017} show that they could use Cellulo to simulate the motion of planets orbiting around the sun or collectively model the energy levels of atoms in the three states of matter. In these cases, Cellulo takes on a more active role in the learning process. By simulating objects in the plane, Cellulo aligns with Ishii and Ullmer's \cite{ishii_tangible_1997} definition of tangible user interfaces which seamlessly bridge the divide between the physical environment and digital cyberspaces.

\section*{Related Works}

\subsection*{Collaborative Remote Learning}
Previous studies have shown that collaborative remote learning courses can deliver high student satisfaction. In order to successfully run remote learning courses, Morgan et al. \cite{morgan_best_2020} recommend giving students collaborative, discovery-based tasks instead of independent worksheets, to allow them to function as active learners rather than passive receivers of knowledge. Given the benefits of tangible interfaces on both exploration and collaboration \cite{schneider_benefits_2011}, low-cost tangible user interfaces (TUIs) like Cellulo \cite{ozgur_haptic-enabled_2017} therefore seem to be prime candidates for remote learning activities. Acknowledging the gap in research on tangible interfaces which function in remote settings, there is broad scope to design tangibles which enhance collaboration skills remotely.

Collaborative learning involves developing and arbitrating shared meanings over the course of a joint task \cite{dillenbourg_what_1999}. To achieve this, participants have to monitor each others' activity within a given problem domain \cite{kianzad_collaborating_2019} and eventually reach some consensus over reaching some mutual goal. One crucial aspect of the collaborative process involves taking on various roles to further explicate the problem space. Dillenbourg \cite{dillenbourg_what_1999} suggests that the \say{collaboration contract} of a task should include a scenario which contains various roles for users. The doling out of roles generates interaction which in turn allows for learning to take place. In \cite{kianzad_collaborating_2019}, the virtual electrostatic lab activity allocates different roles to each participant holding a haptic force-feedback enabled pen. One user is designated the role of placing charges, while the other manipulates the main charge; these roles can be swapped by exchanging the magic pens physically, alluding to the malleable and dynamic nature of role-playing in a collaborative activity. In our computer-supported suite of activities, we aim to dole out roles in a less prescriptive manner by simply giving different visual perspectives to users. As each user physically moves their Cellulo around the various regions on the map, a corresponding visual perspective of the region on the computer screen shifts accordingly. By essentially giving each user a robotic and virtual avatar, they are free to assume roles such as leader or follower at their own pace in an organic manner.

\subsection*{Tangible User Interfaces for Learning}
Tangible robots present opportunities to stimulate collaborative learning. In a project utilising tabletop robots, Neto et al. \cite{neto_using_2020} designed an inclusive learning experience by using force-feedback equipped hand-held Cellulo robots to help pupils learn shapes and letters. Tabletop robots were found to be a successful collaborative tool, with all learners (n=20) collaborating in the experiment with high engagement. Although several of the students had visual impairments, visual acuity of the child did not have an impact on performance of the pair in identifying shapes and letters \cite{neto_using_2020}. This suggests that platforms such as Cellulo \cite{ozgur_cellulo_2017}, RoboGraphics \cite{guinness_robographics_2019} and the Haptic Video Player \cite{guinness_haptic_2018} can be useful for creating collaborative activities for students with visual impairments. Furthermore, the forms of collaboration engineered using TUIs can be dependent on the interaction design of the experiment. In \cite{chibaudel_if_2020}, the authors set up a spatial collaborative treasure hunt task in pairs whereby one person plays the role of an explorer in a large room containing the treasure and another person guides them remotely. A Cellulo robot was used by the guide in two modalities: an active condition where the robot tracked the explorer's location in the room in real time, and a passive condition where the robot was merely a passive figurine representing the explorer whose position the guide should adjust accordingly. The authors found that the active condition could be suitable for collaboration involving directional vocabulary for users with vision impairments \cite{chibaudel_if_2020}. However, the passive condition showed promise in generating a more constrained collaborative environment for users with better verbal communication and vision. The research findings therefore suggest that not only are tangibles a useful tool for fostering collaboration, but that the interaction design of TUIs can in fact alter and transform the resulting collaborative environment.

Limitations of previous studies include either the exclusion of haptic feedback in activities or a lack of multiple interesting modes of haptic feedback if haptic feedback is present in the activity. This means that little is known about what kinds of force-feedback can improve dimensions of collaborative learning. Most studies also only allow for in-person use of tangibles for collaboration, rather than remote collaboration. While Neto et al. \cite{neto_using_2020} make use of haptic feedback in their tabletop robots project, the activity design does not explore different modes or designs of haptic feedback, and functions only in an in-person context. Conversely, \cite{chibaudel_if_2020} does explore interesting interaction designs using tangible interfaces and the impact of this on collaboration, but it does not employ haptic force-feedback and again functions solely in an in-person context. Our system aims to address this gap in the research by developing collaborative remote activities which use haptic feedback, and by creating multiple haptic interaction designs of interest to stimulate collaborative learning.

\subsection*{Haptic Interfaces}
Haptic technologies have been designed to solve problems in myriad fields such as virtual and augmented reality, medicine, rehabilitation and education \cite{noauthor_haptipedia_nodate}. In particular, there is an emergent body of research on haptic feedback in learning environments. While it remains unclear whether haptic feedback can directly improve learning, there appear to be alternate mechanisms through which haptic feedback can improve learning. In \cite{khodr_allohaptic_2020}, students participated in a task whereby they had to \say{feel the curve} of linear functions using force-feedback enabled Cellulo robots. The authors found no significant difference in learning gain between experimental groups with and without haptic feedback \cite{khodr_allohaptic_2020}. However, the presence of haptic feedback significantly increased robot use, which in turn is correlated with learning. This suggests that the mechanisms through which haptic feedback affects learning processes may be diverse and indirect and should be elucidated through further research. In a related study, Johal et al. \cite{johal_learning_2020} utilised haptic feedback-enabled robots to render the borders of geometric shapes in an activity involving abstract geometric concepts such as reflective symmetry. The goal was to remedy common misconceptions students harbour when learning geometry, such as mixing up translations and reflections. The authors found that using haptic robots was about as effective as traditional paper-based methods in dispelling common misconceptions about geometric rules \cite{johal_learning_2020}. This suggests that the haptic modality can indeed be successfully used to render 2D shapes, and can help children gain abstract geometric knowledge of these shapes. These recent studies give valuable insight on how haptic feedback can function as a mode of interaction for learning.

In both of the aforementioned studies, haptic feedback is used to model specific mathematical contexts: linear function gradients in the former and geometric shapes in the latter. As a point of difference, our studies aims to broaden the scope of haptic design and engineer haptic feedback for collaborative communication rather than solely for rendering content and shapes directly from the subject matter. We aim for these interactions to be possible in an online classroom environment rather than solely in a traditional in-person classroom as these previous projects have implemented. It also remains to be investigated if haptic feedback which is targeted to improve collaboration can also elicit improvements in learning gain, which our study aims to address.

Studies into haptic-enabled online learning systems are ongoing. Kianzad et al. \cite{kianzad_collaborating_2019} propose a set of collaborative activities which can be completed in-person or remotely, using haptic force-feedback enabled pens. These Magic Pens carry the advantage of enhanced mobility and a lack of restrictions on the size of the 2D planar workspace. In one activity - the virtual jigsaw puzzle - the pens are used to model virtual constraints, as the users cannot pass the GUI environment and are not allowed to pick up a partner's part \cite{kianzad_collaborating_2019}. They also allow for active guidance: over the course of an activity, an instructor can apply torque in a particular direction in order to guide students towards an area of interest. These few features allude to the range of kinesthetic expressivity which can be achieved using haptic-enabled interfaces. Kianzad et al. \cite{kianzad_collaborating_2019} characterise haptic communication channels as unobtrusive and capable of decluttering GUI-based constraints, thereby lowering the cognitive load of an activity. This study bears similarities to our project in terms of being haptic-enabled and offering online learning options, but it focuses on modelling individual constraints and providing guidance from an instructor through the haptic channel. In contrast, our study focuses on haptic collaboration between participants - in particular, in building consensus and information pooling between partners. Furthermore, our case study uses Cellulo robots which allows for the designation of active and passive modes in activities (iie. where one user can allow another to take remote control of the others' Cellulo) - this is not possible using magic pens.

\section*{Learning Scenario Design} 
In conjunction with the Cellulo robot placed on the map, a virtual dimension was added to the activity in the form of a computer-based online application. This application linked up with the Cellulo robot to display real-time information about the organelle which the robot was current placed upon, creating a seamless connection between the physical and digital learning interfaces (Figure \ref{fig1}). The online application supplied information which the simplified paper map could not: it displayed a 4K view of the organelle under inspection and detailed information of the biochemical processes which occur there. As the user moved the robot around the various organelles on the physical map - for instance, the nucleus or the mitochondria - the application updated with a view of the organelle brought into scope. In most TUIs, there is a metaphorical or symbolic link between the tangible interface and its learning context \cite{celentano_metaphors_2014}. In this scenario, the Cellulo robot was framed as an exploratory device which symbolised the user's journey through the cell. 

The exploratory task as for objectives to introduce the structure of the cell as well as the main characteristics of some organelles (e.g. nucleus, golgi  apparatus,  mitochondrion, lysosome, cytosol). 
The learning content was derived from a course for first year bachelor students in Biology.
The exploration was semi-guided and a series of tasks we proposed to the participants to explore the cell.
Examples of tasks were: "Locate the control centre of the cell", and "Pinpoint the organelle with digestive enzymes". 
Participants were free to do these tasks in the order they wanted and to tick them when they thought they accomplished these tasks. They could do it individually or not.
The participants then moved on to the quiz, finishing the activity. 

The post-activity quiz required participants to agree on their answer before being allowed to submit it, using the Cellulo robots. 
They were give the following list of questions: 
\begin{enumerate}[nolistsep,noitemsep]
    \item\textit{ Which organelle contains most of the cell's genetic material?}
    \item \textit{To which organelle would a protein arrive at to be packaged for export from the cell?}
    \item \textit{Where is chemical energy or ATP (adenosine triphosphate) produced?}
    \item \textit{Which organelle is capable of destroying the cell?}
    \item \textit{What is the space in which all the organelles reside?}
\end{enumerate}
In order to answer, the participants needed to individually place their robot on the organelle which they thought corresponded to the answer. But in order to be able to submit their responses, they had to discuss their choices and agree.

\section*{Interaction Design}
Two haptic modes were designed using the Cellulo robot API  \cite{ozgur_haptic-enabled_2017}. These modes were toggled in a mutually exclusive manner, to control for their effect. They were deployed for the whole duration of the learning activity (including the quiz) to investigate their effect on collaborative online learning. The two modes are described below.

\begin{figure}[t]
    \centering
    \subfloat[Co-location]{
      \includegraphics[width=.45\textwidth]{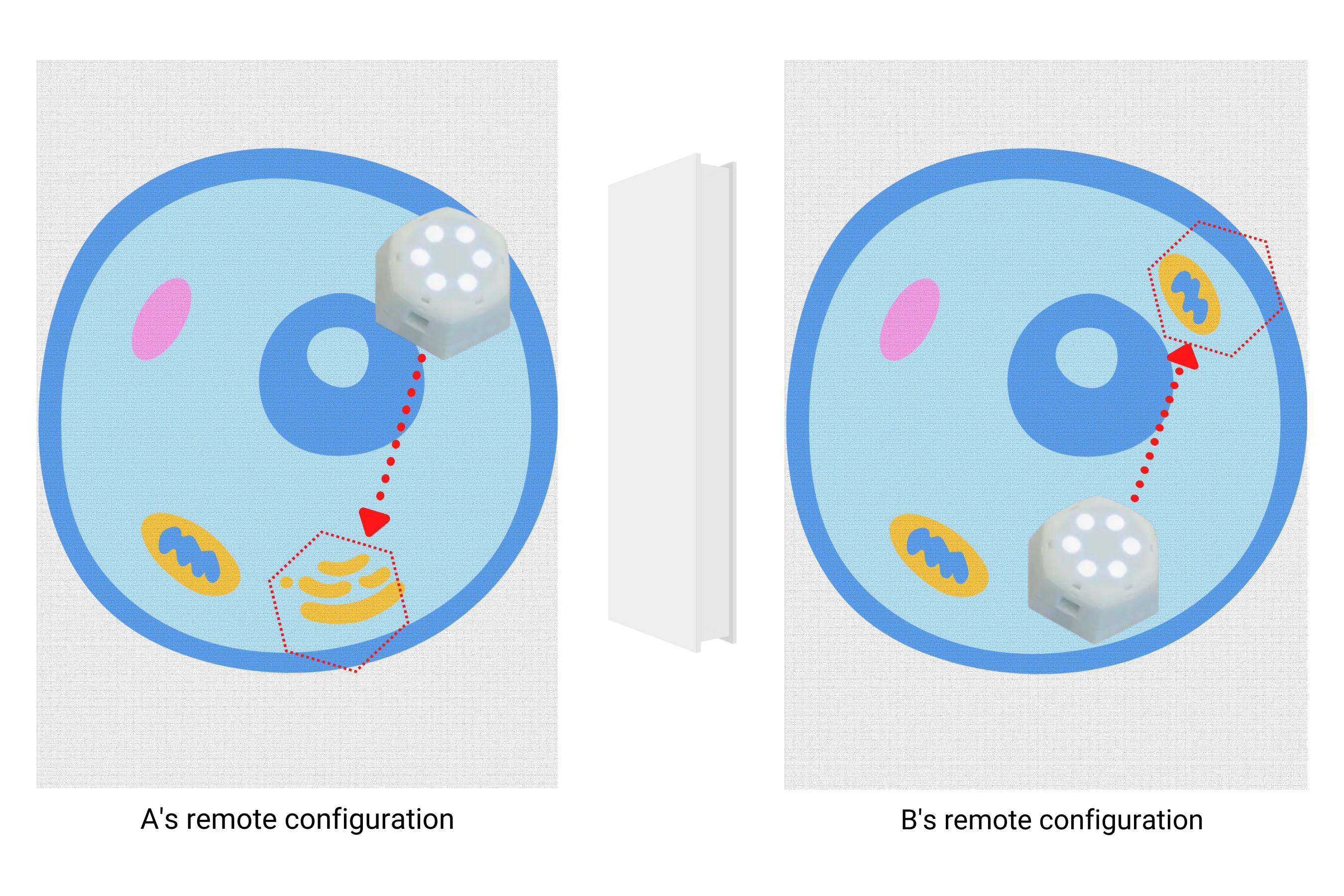}}
    \subfloat[Consensus]{
      \includegraphics[width=.45\textwidth]{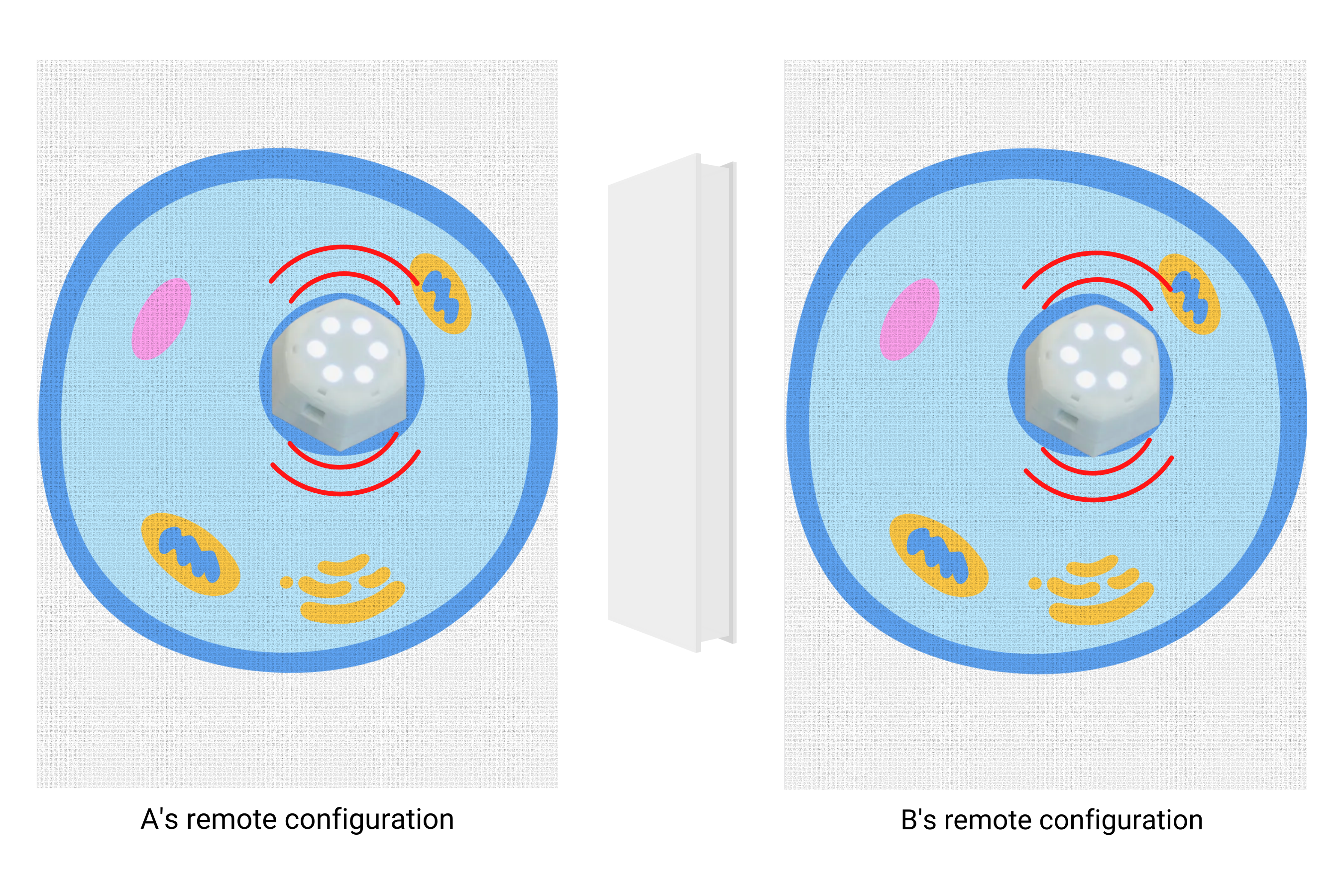}}
    \caption{(a) co-location: Two remotely collaborating participants (A and B) are situated on different organelles of the cell. Exhibited in red, force-feedback pulls the users towards the location of the other. (b) consensus: Two remotely collaborating participants (A and B) are situated on the same organelle (the nucleus). Exhibited in red, the robots emit a gentle vibration as they are in agreement on the same organelle}
    \label{colocation_xy}
\end{figure}

\textbf{Haptic co-location:} When one participant moves one Cellulo to a certain part of the cell map in order to explore the organelle there, the other Cellulo is pulled in that direction as well, modelling the presence of an invisible elastic band binding the two robots together (Figure \ref{colocation_xy} (a)). This feature gives a haptic directional indication of the location of the partner's robot. This is designed to stimulate information pooling, the process of amassing information to find a solution \cite{meier_rating_2007}. When in person, individuals often point or gesture towards information of interest-the haptic co-location feature aimed to substitute this gesture in the remote setting.

\textbf{Haptic consensus:} When two participants have two Cellulo located in the same organelle on their respective maps, a generalised vibration is remotely emitted through the robots (Figure \ref{colocation_xy} (b)). This gives the users a haptic indication that spatial consensus has been reached, thus offering an alternate channel of communication to verbal/visual cues. By providing acknowledgement of consensus in the haptic domain, we also avoid cluttering the GUI with excessive visual cues \cite{kianzad_collaborating_2019}. In-person communication often involves the passing of non-verbal cues to convey agreement, such as nodding or performing a "thumbs up" gesture. The haptic consensus modality aims to mirror these gestures and provide an alternate communication channel to the audio call.

These two haptic modes - \textbf{co-location} and \textbf{consensus} - formed the experimental conditions for comparing the effects of haptic feedback on collaborative learning.

\section*{Case Study}
 \subsection*{Participants  and Study Design}
 4 participants were recruited, consisting of 2 males and 2 females (Mean age: 24.25). The participants were recruited through social media channels. The sample size was greatly limited by the COVID-19 restrictions that were in place at the time. The participants were not reimbursed in any manner for participating in the experiment.

An ethics application under the negligible risk category was submitted to the local university ethics committee. Unconditional approval for the experiment was granted on the 20th of April 2021. 

The 4 participants were grouped randomly into pairs to complete the Cellulo Online collaborative activity, with one pair completing the activity under the haptic \textbf{co-location} mode, while the other pair underwent the haptic \textbf{consensus} mode. For both of the modes, the feature was enabled over the entirety of the experiment including the final quiz. Barring individual differences, this design controlled all other variables in a bid to investigate the differing effects of the haptic modes on collaboration.

\subsection*{Material and Method}
For the two dyads, the participants were placed in two different rooms. Each with the same set of tools.
Each participant was provided a Cellulo robot placed on a printed cell map connected to a computer-based application.
The printed map was a simplified pictorial A4 map of a biological cell and was designed with a microdot pattern, upon which a Cellulo robot could localise itself \cite{ozgur_cellulo_2017}.  It contained the following points of interest:  mitochondria, a lysosome, nucleus,  a golgi body and the cytosol (which is not an organelle but a region of interest).
It should be noted that each Cellulo robot was connected to a separate tablet via Bluetooth in order to transfer its location to the online server. The participants themselves did not interact with the tablets but only had a visual on a computer screen. 
They were provided information about the organelles on a computer screen and could only hear their partner through a Zoom audio call.

The participants began by completing the exploratory phase of the activity, which prompted them to tick off a set of tasks bidding them to explore the contents of the cell.
The participants then moved on to the quiz, finishing the activity. The Zoom audio for each session/pair was recorded for transcription purposes. The position and orientation of the robots on the sheet of paper were also recorded ($x$, $y$, $theta$) for the entirely of each session, in addition to the changes of zones/organelles on which the robots were.

\subsection*{Evaluation}
In the co-location team, the pull of the haptic led to the participants treating the robots as a shared resource. They recognised that it was expedient for them to move together as a pair in the same direction, rather than separating the robots to parallelise the activity through individual work. For instance, they took turns controlling the shared resource, deploying phrases such as \say{Okay, your turn} and \say{Do you want to do one (sic)}, in reference to visiting an organelle on the map. They preferred this over individual divergent exploration. At one stage, the participant states \say{I'll drag \textit{them} [the pair of robots] to random places [organelles]}, indicating that he understands that both robots move together rather than separately. In this sense the participants were interdependent on each other - they could only achieve success by moving as a team. Indeed, the pair visited each organelle in complete tandem (without any external direction from the research team to do so); rather, this was an inference they formed from the haptic feedback. The co-location team took 2:31 minutes to complete the quiz, and the word count of their transcript was 723. They scored $5/5$ for the collaborative quiz.

\begin{figure}[h!]
    \centering
    \subfloat[$x$ position]{
      \includegraphics[width=0.77\textwidth]{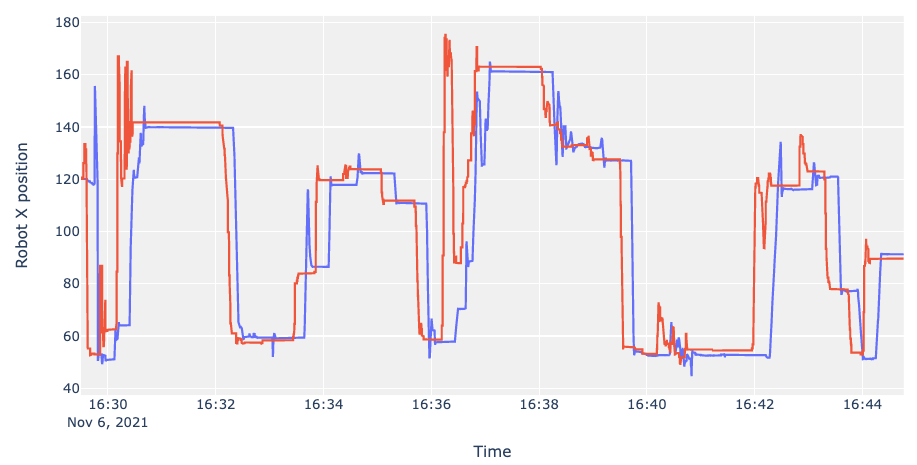}
    }\\
    \subfloat[$y$ position]{
      \includegraphics[width=0.85\textwidth]{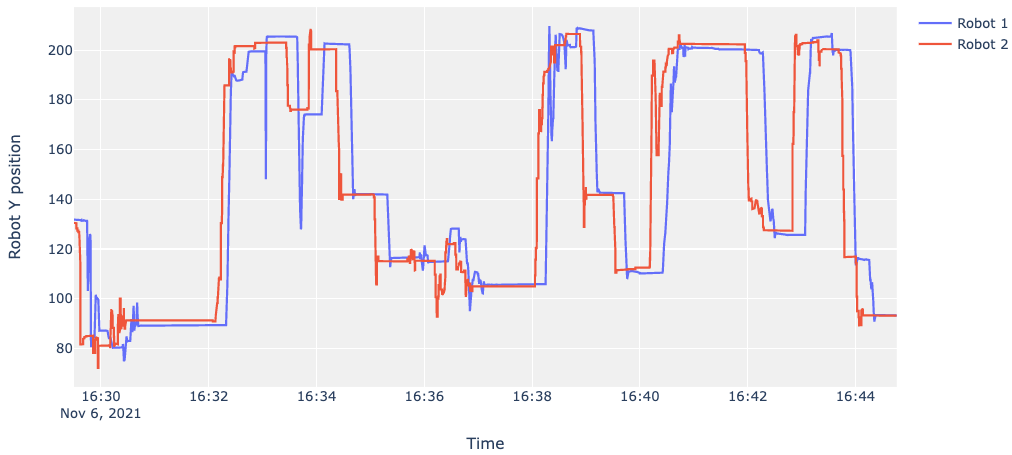}
    }
    \caption{The $x$ and $y$ positions of each robot in the \textbf{co-location} team plotted against
        time, for the duration of the experiment (exploratory activity and quiz).}
    \label{colocation_xy}
\end{figure}

This is well-supported by the robot location data shown in Figure \ref{colocation_xy}, which plots the $x$ and $y$ co-ordinates respectively of the robots operated by each participant in the pair over the experiment. 


Pearson correlations of $0.77$ and $0.82$ were observed for the $x$ and $y$ directions respectively ($p < 0.01$) as per Figure \ref{corr} (a). These strong positive correlations highlight that the robots (and by extension the participants) moved closely together, preferring to converge on locations together than going separate ways.

In contrast, the consensus team tended to act more as individuals than as a singleton. At the end of the exploratory activity, both teams independently came to the conclusion to do a short re-tour of the cell map before beginning the quiz. The co-location team did this in tandem, interacting as they reviewed their knowledge - for instance, they asked each other \say{That's the golgi body?... What does it do? Packaging warehouse - helps process and package proteins and lipids.}. In contrast, the consensus team did a re-tour of the cell map individually, stating \say{I guess we don't have to do it in tandem... We can probably just do it on our own quickly}. This could be attributed to the organisational burden of co-ordinating each others' robots: in the absence of the haptic pulling the pair together, they deploy directional phrases such as \say{Shall we go clockwise} to align their movement. The consensus team completed the quiz in 1:35 minutes and the word count of their full transcript was 403 - both significantly lower than that of the co-location team, lending weight to the notion that greater individual exploration led to less overall interaction. Despite this, the consensus team also scored $5/5$ for the collaborative quiz. Overall, the transcripts suggest a greater degree of collectivism was shown by the co-location group, with individualism exhibited more strongly by the consensus team.

\begin{figure}[h!]
    \centering
    \subfloat[$x$ position]{
      \includegraphics[width=0.73\textwidth]{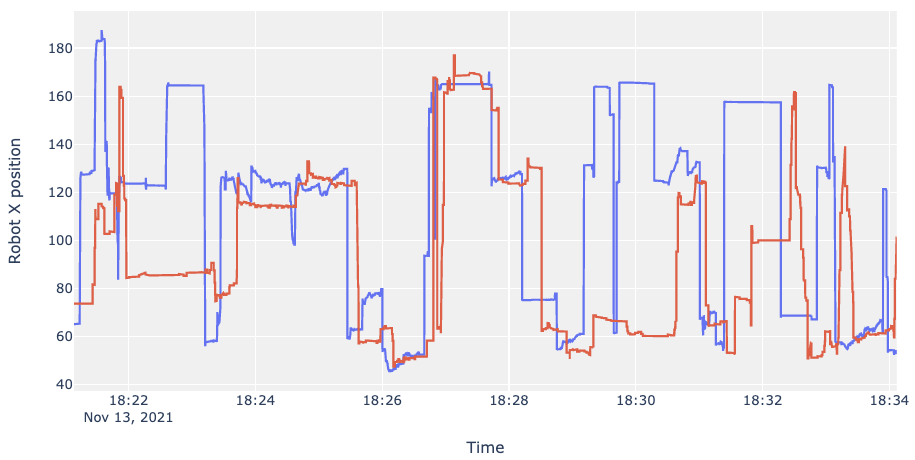}
    }\\
    \subfloat[$y$ position]{
      \includegraphics[width=0.85\textwidth]{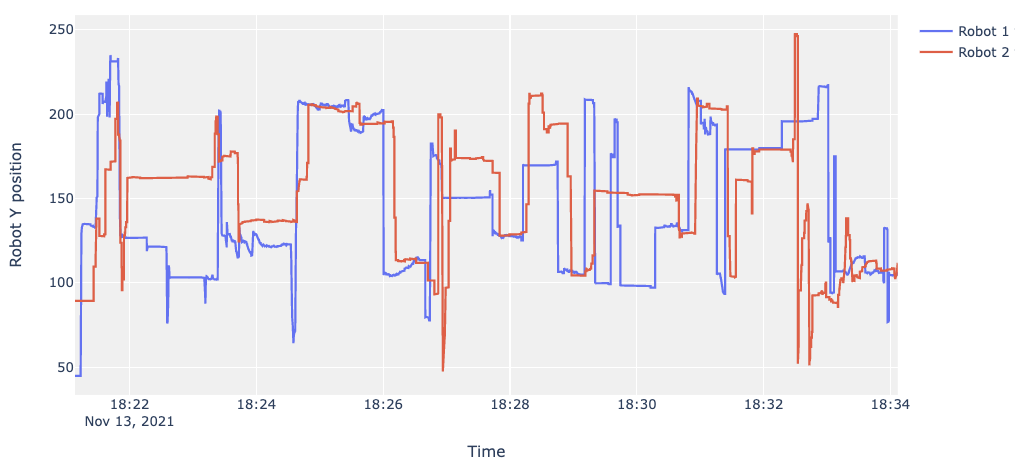}
    }
    \caption{The $x$ and $y$ positions of each robot in the \textbf{consensus} team plotted against
        time, for the  duration of the experiment (exploratory activity and quiz).}
    \label{consensus_xy}
\end{figure}

This disparity is reinforced by the robot location data of the consensus group, with $x$ and $y$ correlations of 0.31 and 0.34 observed respectively ($p < 0.01$). Figure \ref{consensus_xy} demonstrates that while there was certainly some level of mirroring between the movement of the two robots particularly in the window of time between 18:24 and 18:26, this cohesion deteriorates as the experiment goes on. This highlights the increased individualism of the consensus group.

\begin{figure}[h!]
    \centering
     \subfloat[co-location team]{
      \includegraphics[width=0.45\textwidth]{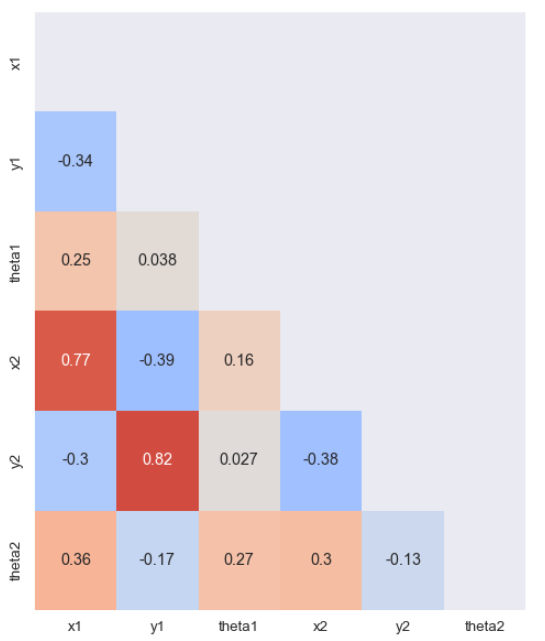}
    }
    \subfloat[consensus team]{
      \includegraphics[width=0.48\textwidth]{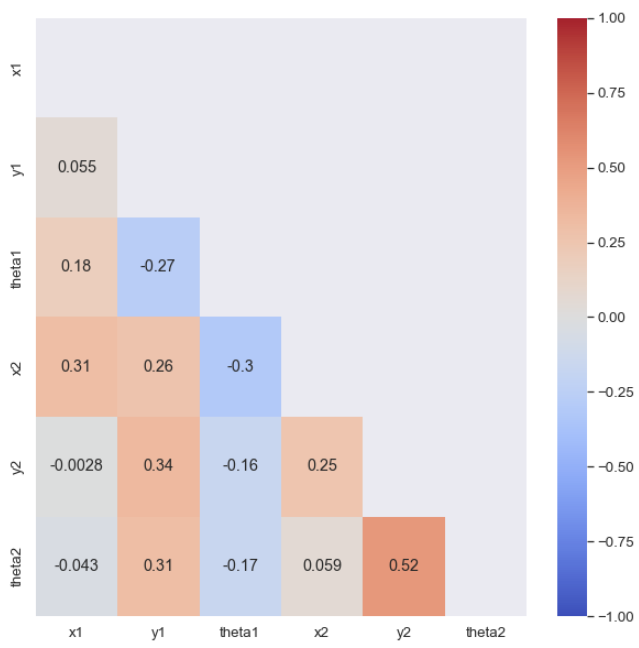}
    }
    \caption{Pearson correlations between each pair of dimensions over the  duration of the experiment (exploratory activity and quiz). For the two pairs of robots ($x1$: the x position of robot 1)}
    \label{corr}
\end{figure}

\section*{Conclusion}
In conclusion, our paper introduces a novel system designed to facilitate haptic communication for collaborative remote learning through tangible robots. We have outlined the system architecture and proposed a case study to assess its efficacy. The results highlight distinct collaborative patterns in two groups—one emphasizing co-location haptic interactions and the other favoring consensus-type haptic engagements.

While acknowledging the study's limitation due to a small participant pool, these preliminary findings serve as a crucial foundation for future endeavors. The case study not only allowed us to validate our setup and metrics but also paved the way for forthcoming large-scale experiments. Our upcoming investigations will delve into leader-follower behaviors and employ the \citep{meier_rating_2007} annotation grid to analyze participant discourse in collaborative learning settings.

Looking ahead, we aim to enhance the complexity of learning content and incorporate pre-tests to measure learning gain disparities. In essence, this work marks an initial stride in exploring haptics as a collaborative communication channel in the realm of remote learning. As we move forward, we envision our research evolving, contributing valuable insights to the intersection of haptic technology and collaborative education.

\section*{Acknowledgments}
We thank the CHILI Lab at EPFL for landing us the Cellulo robots and the NCCR Robotics for supporting this project. 

\nolinenumbers

\bibliography{library}


\end{document}